\documentclass[journal]{IEEEtran}
\usepackage{amsmath}
\usepackage{amsfonts}
\usepackage{color}
\usepackage{bm}
\usepackage{graphicx}
\usepackage{algorithm,algorithmic}
\usepackage{multirow}
\usepackage{cite}
\usepackage[table,xcdraw]{xcolor}

\usepackage{amsmath}
\usepackage{amssymb}
\usepackage{amsthm}
\usepackage{subfigure}
\usepackage{textcomp}
\usepackage{xcolor}
\usepackage{booktabs}
\usepackage{url}

\begin{document}
%
\title{Laxity-Aware Scalable Reinforcement Learning for HVAC Control}
%
%
%

\author{Ruohong Liu, 
 Yuxin Pan, 
 and Yize Chen$^1$
         \thanks{Ruohong Liu and Yize Chen are with the Artificial Intelligence Thrust, Hong Kong University of Science and Technology (Guangzhou), email: rliu519@connect.hkust-gz.edu.cn, yizechen@ust.hk. Yuxin Pan is with the Division of Emerging Interdisciplinary Areas under IPO,  Hong Kong University of Science and Technolog, email: yuxin.pan@connect.ust.hk.}}

\markboth{In submission}%
{Shell \MakeLowercase{\textit{et al.}}: Bare Demo of IEEEtran.cls for IEEE Journals}
%



\maketitle

\begin{abstract}
Demand flexibility plays a vital role in maintaining grid balance, reducing peak demand, and saving customers' energy bills. Given their highly shiftable load and significant contribution to a building's energy consumption, Heating, Ventilation, and Air Conditioning (HVAC) systems can provide valuable demand flexibility to the power systems by adjusting their energy consumption in response to electricity price and power system needs. To exploit this flexibility in both operation time and power, it is imperative to accurately model and aggregate the load flexibility of a large population of HVAC systems as well as designing effective control algorithms. In this paper, we tackle the curse of dimensionality issue in modeling and control by utilizing the concept of laxity to quantify the emergency level of each HVAC operation request. We further propose a two-level approach to address energy optimization for a large population of HVAC systems. The lower level involves an aggregator to aggregate HVAC load laxity information and use least-laxity-first (LLF) rule to allocate real-time power for individual HVAC systems based on the controller's total power. Due to the complex and uncertain nature of HVAC systems, we levrage a reinforcement learning (RL)-based controller to schedule the total power based on the aggregated laxity information and electricity price. We evaluate the temperature control and energy cost saving performance of a large-scale group of HVAC systems in both single-zone and multi-zone scenarios, under varying climate and electricity market conditions. The experiment results indicate that proposed approach outperforms the centralized methods in the majority of test scenarios, and performs comparably to model-based method in some scenarios.
\end{abstract}


%
\IEEEpeerreviewmaketitle

\section{Introduction}
Renewable energy sectors have seen fast growth over the recent years  worldwide\cite{iea2018market}. Yet the intermittency and stochasticity of renewable resources have posed greater operating challenge for modern power networks~\cite{zakaria2020uncertainty}. For instance, in 2020 it is reported by California grid (CAISO) that $33\%$ of electricity was generated from renewable sources, yet certain periods saw renewables curtailment rate as high as $20-30\%$. Without proper actionable solutions, such issue is likely to grow non-linearly \cite{spangher2020augmenting}. On the other hand, the controllable part of electricity demand provide the flexibility with regards to time and location, which is becoming a viable approach to better accommodate uncertain renewables. 
Buildings account for about 40\% of total end-use energy consumption, while the largest energy consumption within building comes from the HVAC system~\cite{chen2018measures}. Therefore, to address the uncertainties arising from both generation  and demand side, deferrable HVAC systems hold huge potential to provide demand flexibility by adjusting their power rate and operation schedule. 

However, effectively utilizing demand flexibility can be challenging. Such challenges come from i). quantifying HVAC flexibility and ii). designing efficient  energy management strategies based on the flexibility and the state information \cite{li2021energy}. Previous attempts use exact load models to realize power system operation goals. Such flexible resources can help accommodate the variability and uncertainty of Renewable Energy Sources (RES)~\cite{contreras2016decentralized, vandoorn2010active}, participate in demand response~\cite{lu2020operation},  reduce energy costs \cite{shi2021end,chen2021smoothed}, and provide ancillary services~\cite{razmara2017building, li2018assessing}.

To take advantage of demand flexibility by hedging against HVAC load stochasticity, it is necessary to firstly define and measure the flexibility of loads. For individual appliance, flexibility is related to the consumer’s preferences and the constraints of appliances. Instead of knowing the exact load and energy consumer's information, \cite{sajjad2016definitions} measures the flexibility of individual appliance by the maximum time an appliance can be postponed without affecting the consumer's comfort. For Electric Vehicle (EV) charging scheduling, the laxity is introduced to measure emergency level of EV's charging need \cite{kwon2021efficient, chen2021smoothed}, and achieves near-optimal results in simulation.


%

In this paper, we develop both modeling and learning-based optimization for HVAC systems through laxity representation. We leverage the fact that efficient aggregation is achievable by utilizing the emergency information of each HVAC control task. We model the HVAC loads as operational requests with differing \emph{laxity} levels based on user-defined setpoint temperatures while subject to zone temperature and HVAC operational range constraints. Essentially, a lower laxity value indicates a more urgent request. Such representation allows the system operator to dispatch power based on the laxity value of the HVAC operation requests rather than solving a complex optimization problem with full state and dynamics information involved. Our design is also privacy-preserving and efficienct, as HVAC systems' information, including heat transfer dynamics model parameters, temperature constraints, and users' thermal preferences are not revealed to the aggregator or controller. 
The aggregator serves as an intermediary to aggregate buildings' flexibility resources while coordinating with the controller to schedule the operation of HVAC systems under time-varying electricity prices~\cite{burger2017review}.  
\begin{figure}[htbp]
\centerline{\includegraphics[width=0.45\textwidth]{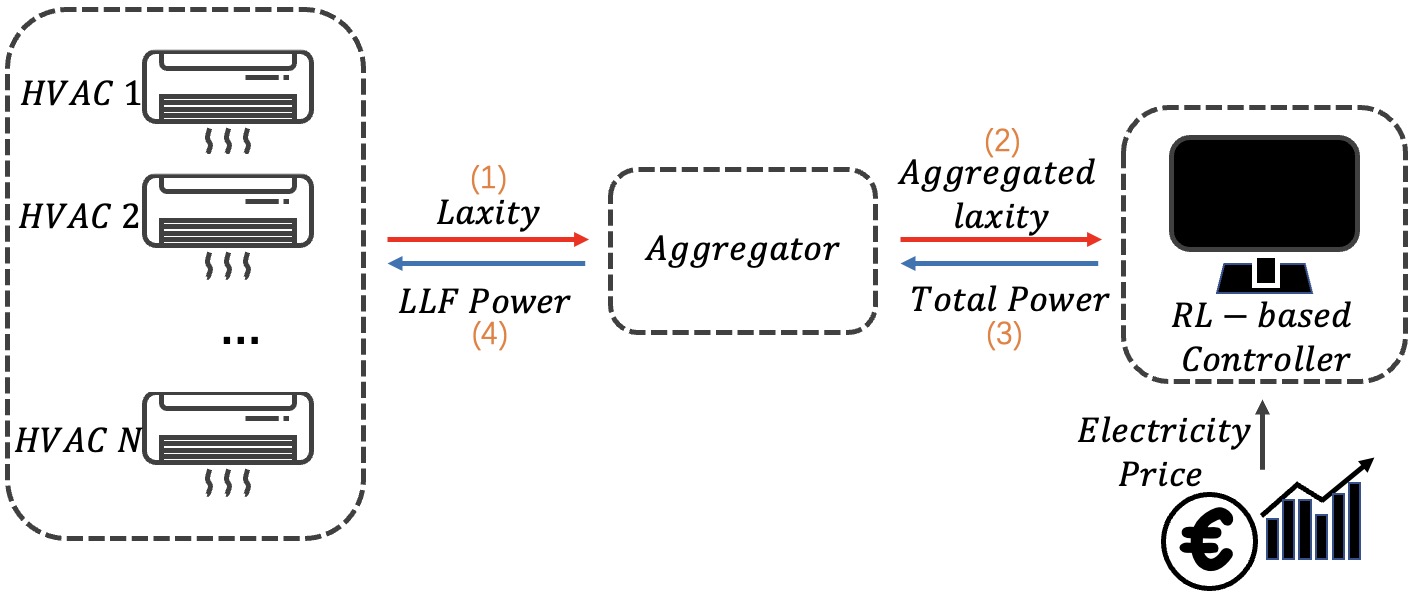}}
\caption{The workflow of our proposed framework (1) HVAC systems report laxity to aggregator; (2) aggregator reports aggregated laxity information to the controller; (3) given aggregated laxity and electricity price information, the RL-based controller schedule total power; (4) given total power, the aggregator recovers power for individual HVAC system according to LLF rule.}
\label{framework1}
\end{figure}

In addition to aggregating each HVAC's flexibility via laxity representation, it is also important to develop corresponding mechanism to schedule the power dispatch.  Yet the electricity price is often time-varying and hard to predict, while building termpoal dynamics are complex and often unknown. In consequence, model-based approaches which require detailed model information are not scalable or efficient. Instead of directly using RL to dispatch power for each HVAC, we propose a model-free RL-based controller only using aggregated laxity to schedule the total power of a larger population of HVAC systems given laxity and electricity price information, as illustrated in Fig. \ref{framework1}. Furthermore, to overcome the curse of dimensionality problem that arises with an increasing number of participating HVAC systems \cite{gosavi2009reinforcement}, we introduce the least-laxity-first~(LLF) rule to recover individual HVAC's power given the total power schedule. We show that LLF rule maintains the feasibility of a feasible total power schedule. We also formally show that with state abstraction, we can reduce the state space of the controller while ensuring that RL agent can learn an approximately optimal policy. The benefits of such design are twofold: it greatly narrows down the action space of RL to the aggregated total power for the HVAC cluster;
and such laxity-based representation can help improve the reliability of RL, while also reducing computational complexity during training and inference. 
The major contributions of this paper are listed as follows:
\begin{enumerate}
    \item We propose a novel modeling of laxity for HVAC system based on both task expected duration and user's thermal preference. In order to include the temperature constraints into consideration, we propose the constraint-augmented laxity. 
    \item To efficiently utilize flexibility of a group of HVAC systems, we develop an aggregation framework which only requires HVAC individual laxity, and maps all laxity information to an abstract state. Such aggregated, learning-based controller can efficiently find power dispatch minimizing the electricity bill. 
    \item We develop LLF power dispatching rule and prove that with LLF rule, the aggregator can guarantee a feasible power schedule for each request from a feasible total power schedule only based on its laxity value. 
\end{enumerate}






\section{Literature Review}
\label{review}
To model and quantify demand shifting and load control potential, researchers investigate approaches on measuring the flexibility of building operation requests, especially for HVAC loads and EV loads. \cite{sajjad2016definitions} measures the flexibility of individual appliance by customer’s Acceptable Delay Time (ADT), which measures the maximum allowable delay of control actions without sacrificing consumers’ comfort; while the Appliance Flexibility Index (AFI) measures the adjustable range of appliance operation time. The notion of flexibility envelope is proposed in \cite{cai2021experimental}, which adopts a model-based approach and characterizes flexibility as a three-dimensional tensor spanned by lead time, feasible power levels, and corresponding maximum sustained duration. Yet the calculation of flexibility envelop still is time-consuming, and hard to scale to control of multiple residential buildings~\cite{hekmat2021data}. 

In contrast to fixed-priority scheduling algorithms which only assigns the priority once, LLF is a more effective dynamic-priority scheduling algorithm which is also simple to implement~\cite{leung1982complexity}. 
LLF is further adopted in many related studies in EV charging station scheduling, where laxity is used to measure the emergency level of EV's charging demands, as well as constraints on duration and maximum power \cite{kwon2021efficient, chen2021smoothed}. Recent attemps also consider the machine learning problem under complex and unknown dynamics with laxity involved\cite{wang2019reinforcement, kwon2021efficient}. To the best of our knowledge, laxity is rarely discussed for control of HVAC. While in \cite{alizadeh2013least} laxity is introduced for Thermostatically Controlled Loads~(TCL) control, only an ON/OFF control is discussed without incorporating thermal comfort level or demand response incentives. 


Previous research also work on online scheduling of HVAC by   leveraging HVAC load flexibility~\cite{lilliu2023heat}.
Taking the perspective of modeling load shifting potentials, aggregating flexibility of a group of loads is succinctly treated as stochastic virtual batteries\cite{hao2014aggregate, zhao2017geometric}, and has been applied to various tasks such as establishing scheduling model to maximize customers' total profit \cite{wu2019multi}, and providing ancillary service such as frequency regulation for power system \cite{hughes2016identification}. The energy management system can make sequential turn-on/turn-off decisions for appliances to make a trade-off between users' comfort and energy savings in response to time-varying electricity price \cite{jin2017foresee}. Learning and stochastic optimization framework is also proposed for residential demand response under price incentives \cite{chen2021online, chen2017modeling}.
Though efficient for modeling such aggregated load behaviors, most of them focus on the demand signals tracking tasks \cite{hughes2016identification, alizadeh2013least}. It is yet to explore designing controllers directly using laxity as state representations.

Our work is most closely related to research on leveraging the potential of data-driven approaches for energy management for HVACs and smart buildings. Learned from interactions with the environment, RL algorithms can learn the complex and nonlinear mapping between system states and optimal actions~\cite{mocanu2018line}. RL can control energy storage devices in building clusters to perform energy arbitrage
~\cite{liu2023learning}. Multi-agent RL methods have also been applied to energy optimization problems in buildings, with the aim of meeting user demands while reducing energy consumption \cite{yu2020multi}. 
However, most previous literature use full state representation, and it is computationally expensive to train policies for multiple HVAC systems. It is thus promising to design state representation and abstraction approaches on the laxity for a set of HVAC requests \cite{jiang2018notes}. Meanwhile, it is demonstrated that given a feasible total power schedule and well-designed LLF rule, there always exists a recovered feasible schedule for individual HVAC systems \cite{wang2019reinforcement}. Therefore, our framework aims to reliably optimize power consumption with an RL-based controller, which is ensured to provide a feasible total power schedule taking into account both laxity and external electricity price. 

To achieve efficient online scheduling for multiple HVAC systems, we address the following challenge in this work: 

\emph{How can the controller obtain an efficient operating schedule which is feasible to individual HVAC operation request? }



\section{Problem Formulation}
\label{Problem}
\subsection{HVAC System Model}

\begin{figure}[htbp]
\centerline{\includegraphics[width=0.35\textwidth]{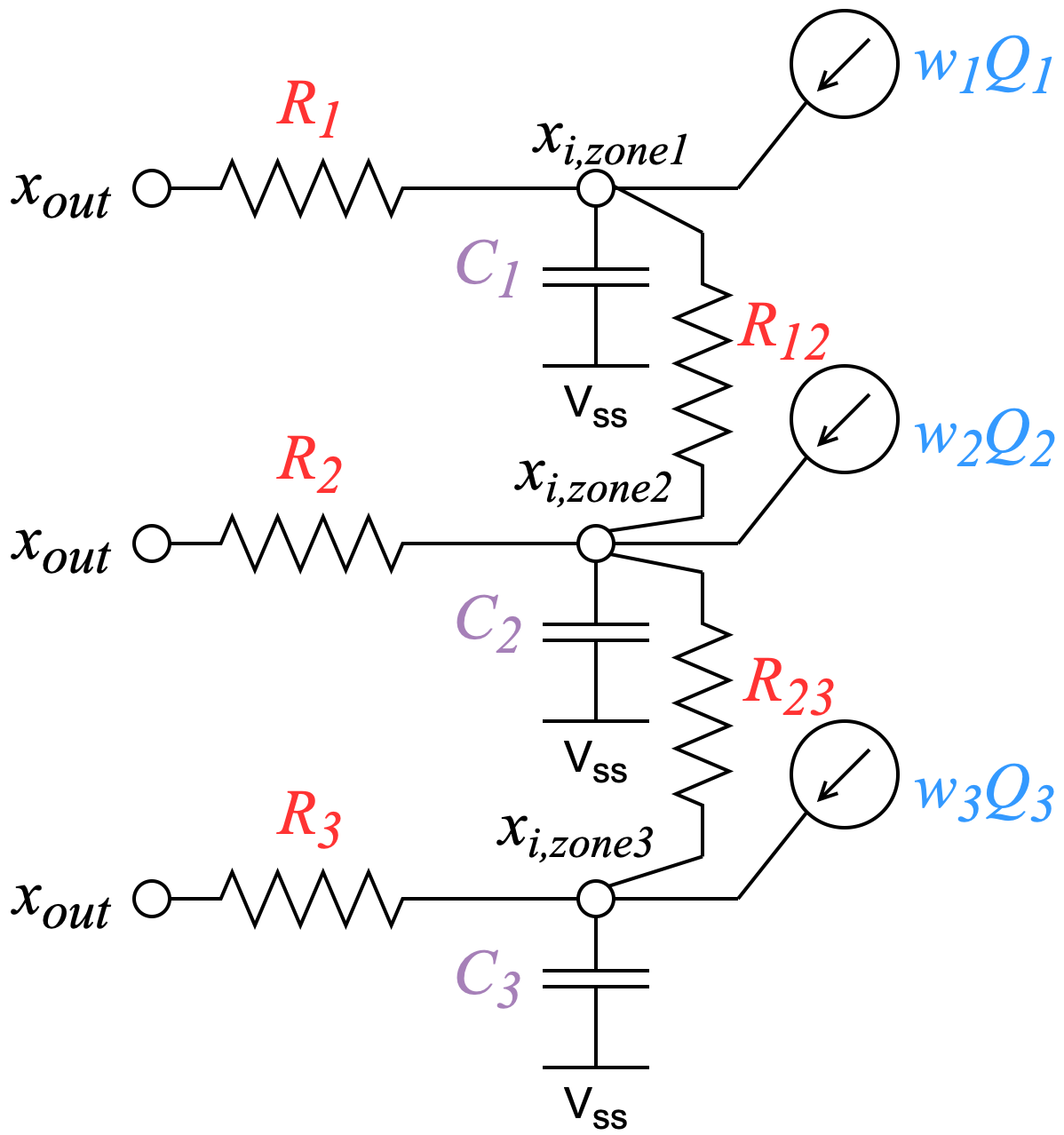}}
\caption{An illustration of  three-zone RC thermal dynamics model structure.}
\label{RC}
\end{figure}

We consider controlling the power dispatch for a set of $N$ HVAC systems, indexed by $i \in \mathcal{N}=\{1,2,3, \cdots\}$, and  we assume each zone in the building is equipped with an HVAC system using the same index. To enable modeling and learning via Markov Decision Process (MDP), we discretize the timestep by $\Delta t, \, t \in \mathcal{T}=\{1,2,3, \cdots\}$. The following thermal resistor-capacitor networks (RC models) represent the general thermal dynamics of buildings \cite{wang2019development}: 
\begin{equation}
\label{state function}
\resizebox{1\linewidth}{!}{$\dot{x}_i [t]=\sum_{j\in\mathcal{M}(i)}\frac{1}{R_{ij} C_{i}}(x_{j}[t]-x_i[t])+\frac{1}{R_i C_i}(x_{out}(t)-x_i[t]) + \frac{w_i}{C_i}u_i[t],$}
\end{equation}
where $x_i[t], \, x_{out}[t]$ denote the indoor temperature of the $i^{th}$ zone and outdoor temperature at the $t^{th}$ time step respectively; $\mathcal{M}(i)$ are the neighbor zones of the $i^{th}$ zone;  $u_i[t]$ is the control action for the $i^{th}$ HVAC system, and without loss of generality, it denotes heating (positive) or cooling (negative) power; $C_i$ is the thermal capacity attached to the $i^{th}$ building; $R_i$ is the thermal resistance between the $i^{th}$ zone and outdoors; $R_{ij}$ is the thermal resistance between the $i^{th}$ zone and its neighbor zone; $w_i$ is a weighting factor which quantifies the efficiency of control input. 
Fig. \ref{RC} shows an example of RC model for a multi-zone building. 

We first develop our proposed method on the single-zone setting with a simplified thermal dynamics model:
$\dot{x}_i [t]=\frac{1}{R_i C_i}(x_{out}[t]-x_i[t]) + \frac{w_i}{C_i}u_i[t],$ 
and will address the multi-zone building case in Section \ref{case study}. The general zone-level dynamics can be formulated as
\begin{equation}
\label{state function}
\dot{x}_i [t]=a_i(x_{out}[t]-x_i[t]) + b_iu_i[t],
\end{equation}
where $a_i$ is the average heat loss rate; $b_i$ denotes the conversion efficiency related to electrical and thermal power.

We assume each building HVAC user has its preferred temperature $x_{i,r}$ and comfortable temperature range:
\begin{equation}
    \underline{x}_{i}\leq x_{i}[t]\leq \bar{x}_{i}; 
\end{equation}
along with limitations on heating/cooling power:
\begin{equation}
\label{equ:action}
    \underline{u}\leq u_i[t] \leq \bar{u};
\end{equation}
where $\underline{u}$ and $\bar{u}$ denote the maximum cooling and heating power, respectively. Without loss of generality, we assume $\underline{u} = -\bar{u}$ for simplified representation.

\vspace{-10pt}\subsection{Optimization Problem}
In this work, we focus on controlling a group of HVAC systems that can take heterogeneous dynamics~\eqref{state function} under time-varying electricity price set by a tariff. The objectives of the HVAC control problem are to minimize the difference between the actual indoor temperature of each building and the corresponding target temperature as well as energy cost. 
\begin{equation}
\begin{aligned} \label{P}\vspace{-10pt}
&\min _{\{u_i[t]\}_{t\in \mathcal{T}}} \quad \sum_{t \in \mathcal{T}} \sum_{i \in \mathcal{N}} \left( c[t] |u_i[t]|+ \left|x_{i}[t]-x_{i,r}\right| \right) \\
s.t. & \quad \;\eqref{state function}-\eqref{equ:action};\\
&\quad \;\sum_{i\in \mathcal{N}} |u_i[t]|\leq P[t], t\in \mathcal{T};\\
&\quad \;\underline{P} \leq P[t]\leq \bar{P}, t\in \mathcal{T};
\end{aligned}
\end{equation}
where $c[t]$ is the time-of-use electricity price, $P[t]$ is the total injection power with limits $\underline{P}$ and $\bar{P}$. To explicitly solve this optimization problem for a cluster of buildings, it is necessary to obtain the exact model knowledge of each HVAC system, including their model parameters and temperature states, and upload such information to an aggregator for processing~\cite{razmara2017building}. Moreover, it is impractical for the HVAC systems, the aggregator  and the controller to communicate complex dynamics and constraints due to privacy concerns or computational complexity \cite{li2021learning}. Therefore, it is crucial to seek efficient feasibility representation and HVAC control strategy while taking both objective and constraints into account.

\vspace{-10pt}
\subsection{HVAC Operation Request}
Before delving into the definition of laxity $l_i$ for each HVAC agent, we first  define the operation requirement of each HVAC system as an HVAC operation request. 
The operator operates each HVAC in each interval. A request is expected to be finished before the expected ending time. Let $\mathcal{C}[t]$ represent a collection of HVAC operation requests to be executed at time step $t$, which is defined as follows: 

\textbf{Definition 1 (Request)} An HVAC operation request is characterized with a quintuple $\mathcal{J}=\left( t_{i,s},t_{i,e},\tau_i,l_i \right)$, where $t_{i,s}$ and $t_{i,e}$ denote the starting time and expected ending time of a request, respectively; $\tau_i$ and $l_i$ denote the \emph{penalty} and \emph{laxity}, respectively. 
Each HVAC system is allowed to send a new request when any of the following conditions is true: 
\begin{itemize}
    \item Current request misses expected ending time, i.e, $t>t_{i,e}$;
    \item Current request is finished, which indicates the indoor temperature reaches the target temperature, i.e., $\left(x_i[t-1]-x_{i,r}\right)\left(x_i[t]-x_{i,r}\right)\leq 0$;
\end{itemize}

\section{Laxity Representation}
\label{laxity representation}
\subsection{Laxity and Penalty Representation}
\begin{figure}[htbp]
\centerline{\includegraphics[width=0.45\textwidth]{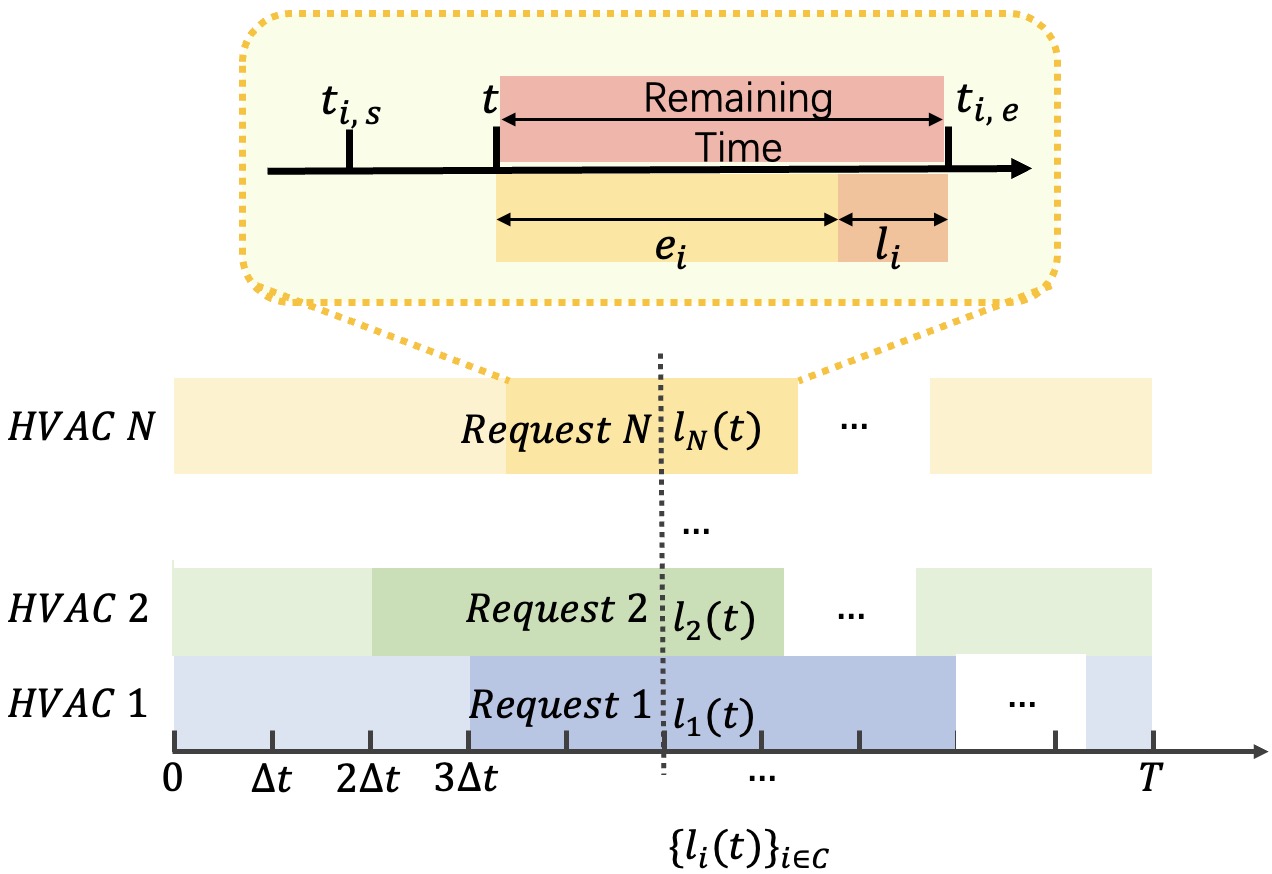}}
\caption{Laxity information reported to the aggregator. After completing its operation request or missing the expected ending time, the HVAC system initializes a new request. At each timestep, the HVAC systems report their respective laxity information to the aggregator.} 
\label{request}
\end{figure}

In this paper, we propose to schedule the operation of HVAC systems considering their urgency level, power constraints, temperature constraints, and users' preferred temperature. However, uploading information of all HVAC systems in real time may lead to privacy concerns and communication burdens. In addition, the heterogeneous information of HVAC systems also makes the scheduling task complex. Hence, in this subsection, we will define the laxity and penalty of an HVAC system, which are two issues to be considered in the scheduling process. Fig. \ref{request} depicts a set of HVAC operation requests and how laxity is defined.

Before detailing the definition of laxity and related variables, we first define an operator as follows: $\zeta(x_1,x_2,u,\lambda_i) = -\frac{1}{a_i\Delta t} \ln \left(\frac{x_1-\frac{b_iu}{a_i}-x_{out}[t]}{x_2-\frac{b_iu}{a_i}-x_{out}[t]}\right)$, where $\lambda_i = (a_i,b_i,\underline{x}_{i},\bar{x}_{i},\underline{u},\bar{u})$ denotes the model parameters of the $i^{th}$ HVAC system.

Next, we introduce the definition of \emph{penalty} to measure the level of violating the comfortable temperature range:

\textbf{Definition 2 (Penalty)} Denote $\tau_i[t]$ as the expected job length to get the current temperature from $x_i[t]$ to the boundary $\underline{x}_{i}$ or $\bar{x}_{i}$. When $x_i[t]\notin [\underline{x}_{i} , \bar{x}_{i}]$. 
$\tau_i[t]$ is calculated  by heating/cooling with maximum heating/cooling power:
\begin{equation}
\tau_i[t]=\left\{
	\begin{aligned}
    &\zeta(\underline{x}_{i},x_i[t],\bar{u},\lambda_i),  \quad \text{if $x_i[t]<\underline{x}_{i}$}\\
    &0, \quad \text{if $\underline{x}_{i}\leq x_i[t]\leq \bar{x}_{i}$}\\
    &\zeta(\bar{x}_{i},x_i[t],\underline{u},\lambda_i),  \quad \text{if $x_i[t]>\bar{x}_{i}$}\\
	\end{aligned}
	\right
 .
\end{equation}

Given the fact that \emph{laxity} is defined over time, while HVAC demand is measured in energy (kWh), we propose to also determine the \emph{minimum heating/cooling time}, which is proportional to the energy demand.

\textbf{Definition 3 (Minimum heating/cooling time)}: Minimum heating/cooling time denotes the minimum time to get the current temperature from $x_i[t]$ to the target temperature $x_{i,r}$ with a maximum heating/cooling rate. It measures the energy demand with the number of time steps.
\begin{equation}
\label{equ:request}
e_{i}[t]=\left\{
	\begin{aligned}
    &\zeta(x_{i,r},x_i[t],\bar{u},\lambda_i),  \quad \text{if $x_i[t]<x_{i,r}$}\\
    &0, \quad \text{if $x_i[t]= x_{i,r}$}\\
    &\zeta(x_{i,r},x_i[t],\underline{u},\lambda_i),  \quad \text{if $x_i[t]>x_{i,r}$}\\
	\end{aligned}
	\right
 .
\end{equation}

The definition of $e_i[t]$ and $\tau_i[t]$ are obtained by discretizing the state function~\eqref{state function}. Since there is a logarithmic function in both of them, we first show in the following proposition that the definition of $e_i[t]$ and $\tau_i[t]$ are always valid.

\textbf{Proposition 1}: Assume the heater/air-conditioner can work to heat/cool under normal conditions of outdoor temperature. Then,  $e_i[t]$ and $\tau_i[t]$ are always real numbers.

\begin{proof}
When the HVAC can work to heat/cool under normal conditions of outdoor temperature, i.e. when $x_i[t]\leq x_{i,r}$, the heater can work to heat the room using $\bar{u}: \dot{x}_i[t]=a_i(x_{out}[t]-x_i[t]) + b_i\bar{u}>0, \forall x_i[t]\leq x_{i,r}$; similarly we have $\dot{x}_i[t]=a_i(x_{out}[t]-x_i[t]) + b_i\underline{u}<0$ when $x_i[t]\geq x_{i,r}$. 
\newline
    1). For $e_i[t]$, if $\dot{x}_i[t]=a_i(x_{out}[t]-x_i[t]) + b_i\bar{u}>0, \forall x_i[t]\leq x_{i,r}$, hence $x_{i,r}-\frac{b_i\bar{u}}{a_i}-x_{out}[t]<0$ and $x_i[t]-\frac{b_i\bar{u}_{i}}{a_i}-x_{out}[t]<0$, $\frac{x_{i,r}-\frac{b_i\bar{u}_{i}}{a_i}-x_{out}[t]}{x_i[t]-\frac{b_i\bar{u}}{a_i}-x_{out}[t]}>0$ holds. Similarly, when $\dot{x}_i[t]=a_i(x_{out}[t]-x_i[t]) + b_i\underline{u}<0, \forall x_i[t]\geq x_{i,r}$, 
    $\frac{x_{i,r}-\frac{b_i\underline{u}}{a_i}-x_{out}[t]}{x_i[t]-\frac{b_i\underline{u}}{a_i}-x_{out}[t]}>0$ holds. 
    By proving the domain of the logarithmic term in the definition of $e_i[t]$ is strictly greater than zero, we can ensure that $e_i[t]$ always returns a real number. 
\newline 2). For $\tau_i[t]$, $\dot{x}_i[t]=a_i(x_{out}[t]-x_i[t] + b_i\bar{u}>0, \forall x_i[t]\leq x_{i,r}$, hence $\underline{x}_{i}-\frac{b_i\bar{u}}{a_i}-x_{out}[t]<0$ and $x_i[t]-\frac{b_i\bar{u}}{a_i}-x_{out}[t]<0$, $\frac{\underline{x}_{i}-\frac{b_i\bar{u}}{a_i}-x_{out}[t]}{x_i[t]-\frac{b_i\bar{u}}{a_i}-x_{out}[t]}>0$ holds. Similarly, when $\dot{x}_i[t]=a_i(x_{out}[t]-x_i[t]) + b_i\underline{u}<0, \forall x_i[t]\geq \bar{x}_{i}$, 
$\frac{\bar{x}_{i}-\frac{b_i\underline{u}}{a_i}-x_{out}[t]}{x_i[t]-\frac{b_i\underline{u}}{a_i}-x_{out}[t]}>0$ holds. By proving the domain of the logarithmic term in the definition of $\tau_i[t]$ is strictly greater than zero, we can ensure that  $\tau_i[t]$ always returns a real number. 
\end{proof} 
\textbf{Definition 4 (Laxity)} The laxity of a request $i\in\mathcal{C}[t]$ at time $t$ is defined as the \emph{minimum heating/cooling time} subtracted from the \emph{remaining request duraction}:

\begin{equation}
\label{equ:laxity}
\ell_i[t]= \left\{
	\begin{aligned}
    &+\infty,  \quad \text{if } t<t_{i,s} \\ 
    &t_{i,e}-t-e_i[t],  \quad \text{if } t_{i,s}\leq t \leq t_{i,e}\\
	\end{aligned}
	\right
 .
\end{equation}

Intuitively, in normal situations, the \emph{laxity} of an HVAC operation request decreases as the \emph{remaining duration time} decreases faster than the \emph{minimum heating/cooling time}. We will provide proof of this relationship and establish sufficient conditions for its validity. 

\textbf{Proposition 2}: Assume  $|x_{out}[t] - x_{out}[t+1]|$ is sufficiently small given an incremental timestep $\Delta t$. Then,  HVAC laxity is monotonically decreasing with respect to the time.

\begin{proof}
     Since $\ell_i[t+1]-\ell_i[t]= (t_{i,e}-[t+1]-e_i[t+1]) - (t_{i,e}-t-e_i[t])=-1+(e_i[t]-e_i[t+1])\leq 0,$ it is sufficient to show $e_i[t]-e_i[t+1]\leq 1.$ We now prove the inequality holds for heating process, while the cooling case is symmetrical. With
\begin{equation}
    \begin{aligned}
        &e_i[t]-e_i[t+1]=\zeta(x_{i,r},x_i[t],\bar{u},\lambda_i)-\zeta(x_{i,r},x_i[t+1],\bar{u},\lambda_i)\\
        &=-\frac{1}{a_i \Delta t} \ln \left(\frac{x_i[t+1]-\frac{b_i\bar{u}}{a_i}-x_{out}[t]}{x[t]-\frac{b_i\bar{u}}{a_i}-x_{out}[t]}\right)\\
        &\resizebox{1\linewidth}{!}{$=-\frac{1}{a_i \Delta t} \ln \left(\frac{e^{-a_i \Delta t}(x_i[t]-\frac{b_i u_i[t]}{a_i}-x_{out}[t])+\frac{b_i u_i[t]}{a_i}-\frac{b_i \bar{u}}{a_i}}{x_i[t]-\frac{b_i\bar{u}}{a_i}-x_{out}[t]}\right),$}
    \end{aligned}
\end{equation}
to show $e_i[t]-e_i[t+1] \leq 1$ for one timestep, that is $ \ln \left(\frac{e^{-a_i \Delta t}(x_i[t]-\frac{b_i u_i[t]}{a_i}-x_{out}[t])+\frac{b_i u_i[t]}{a}-\frac{b_i \bar{u}}{a_i}}{x_i[t]-\frac{b_i\bar{u}}{a_i}-x_{out}[t]}\right) \geq -a_i\Delta t$. Then, we need to show:
$
\small
    \left(\frac{e^{-a_i \Delta t}(x_i[t]-\frac{b_i u_i[t]}{a_i}-x_{out}[t])+\frac{b_i u_i[t]}{a_i}-\frac{b_i \bar{u}}{a_i}}{x_i[t]-\frac{b_i\bar{u}}{a_i}-x_{out}[t]}\right) \geq e^{-a_i\Delta t},
$
namely
\begin{equation}
\label{15}
(1-e^{-a_i \Delta t})\frac{b_i }{a_i}(u_i[t]- \bar{u}) \leq 0.
\end{equation}
Because $u_i[t]\leq \bar{u}$, $1-e^{-a_i \Delta t}>0$, and $\frac{b_i}{a_i}>0$, it is sufficient to shows \eqref{15} holds. 
\end{proof}

\subsection{Constraint-Augmented Laxity}
In section \ref{Problem}, we define $\emph{penalty}$ corresponding to the temperature constraint $\underline x_i\leq x_i \leq \bar x_i$. 
In practice, some deferrable loads already incurs penalty, and it is necessary to assign HVAC systems violating the temperature constraints with higher priority. 
If $\emph{penalty}$ is fed directly into the reward function of reinforcement learning, i.e., as an input to the controller, this does not guarantee those HVAC systems violating temperature constraints can be operated with higher priority. Accordingly, we develop an adapted laxity term~\cite{chen2021smoothed} that guarantees those HVAC systems violating temperature constraints have smaller laxity than those do not violate the constraints, thus they can be operated with higher priority.

\textbf{Definition 7 (Constraint-Augmented Laxity)} 

\begin{equation}
\label{augment}
\ell_i[t]= \left\{
	\begin{aligned}
    &+\infty,  \quad \text{if } t<t_{i,s} \\ 
    &t_{i,e}-t-e_i[t],  \quad \text{if } t_{i,s}\leq t \leq t_{i,e} \land \underline{x}_{i}\leq x_i[t]\leq \bar{x}_{i} \\
    &-\tau_i[t],  \quad \text{if } x_i[t]\leq \underline{x}_{i} \lor x_i[t]\geq \bar{x}_{i}.
	\end{aligned}
	\right
 .
\end{equation}

\section{Laxity-Aware Reinforcement Learning}
\label{methodology}
In this section, we will discuss given laxity information, how to efficiently solve the energy optimization problem defined in equation \eqref{P} through online scheduling and power dispatching for individual HVAC systems.

\subsection{Online Scheduling}
In our HVAC controller design, the aggregator first receives operation requests $\mathcal{C}[t]$ from HVAC systems. Then, it aggregates the laxity information $L[t] = \sum_{i}^{n} \ell_i[t], \forall i\in \mathcal{C}[t]$ and sends it to the controller, which calculates the total power $P[t]$ based on aggregated laxity information and electricity price. Next, the controller schedules the real-time total power, then the aggregator dispatches the power $u_i[t]$ for HVAC systems based on LLF rule. The designed controller is more computationally efficient as it only needs to decide $P[t]$.
Note that we use $\mathcal{C}[t]$ and $\mathcal{C}$ interchangeably hereafter.

Algorithm \ref{A1} details how to utilize the least-laxity-first~(LLF) rule to recover power for HVAC systems from a total power schedule based on 
their laxity. With LLF rule, the HVAC operation request with smaller laxity will have the priority to be operated. If all HVAC operation requests have the probability to be satisfied with a certain total power schedule, then the total power schedule is feasible. The specific definition of a feasible schedule is given as:

\textbf{Definition 5 (Feasible schedule)}: A total power schedule $\textbf{P} = (P[1], P[2], ..., P[T])$  for a set of HVAC operation requests set $\mathcal{C}[t]$ is feasible if it satisfies:
\begin{equation}
    \sum_{i\in \mathcal{C}[t]} |u_i[t]|\leq P[t], t\in \mathcal{T},
\end{equation}
\begin{equation}
    \underline{P} \leq P[t]\leq \bar{P}, t\in \mathcal{T},
\end{equation}
\begin{equation}
    e_i [T] = 0, i\in \mathcal{C}[t].
\end{equation}

With LLF rule described in Algorithm \ref{A1}, we show the aggregator can recover feasible operation schedules for all the requests given a feasible total power schedule $\textbf{P}$:

\textbf{Proposition 3}: If there exists at least one feasible total power schedule $\textbf{P} = (P[1], P[2], ..., P[T])$ for a set of HVAC operation requests $\mathcal{C}[t]$, then the LLF rule can recover a feasible operation schedule for all the request set $\mathcal{C}[t]$.
\begin{proof}
As shown in $\textbf{Proposition 2}$, $\ell_i[t]$ is monotonically decreasing at $t_{i,s}\leq t \leq t_{i,e}$. Then, as $e_i[t] = 0, i\in \mathcal{C}[t]$, $\ell_i[t]=0$ when $t=t_{i,e}$. Therefore, a feasible total power schedule indicates that $\ell_i[t]\geq 0, i\in \mathcal{C}[t], t\in T$. 

Now we consider a random step $t=k$ with total power $P[k]$. The total power $P[k]$ ensures that $\ell_{i}[k+1]\geq 0, i\in \mathcal{C}[t]$, i.e. $P[k] \geq 
n \bar{u}$, where $n = |\{j \in \mathcal{C}[t] : 0 < l_j < 1\}|$, i.e. the number of requests with laxity $0 < l_j < 1$. With LLF rule, these $n$ requests are given higher priority due to their smaller laxity compared to other requests. Then for any other requests $j$,  $\ell_{i}[k+1]\geq \ell_j[k+1]= 0$. 
\end{proof}

In Table. \ref{2requestsexample}, a toy example with two simultaneous HVAC operation requests is presented. Given a feasible total power schedule, two HVAC requests are shown with feasible schedules under the LLF scheduling policy. However, if deviating from LLF rule, e.g., by swapping power at time step 3 (in blue), could result in a negative laxity value, indicating potential failure to meet the deadline (in red).

\begin{table}[]
\caption{HVAC operation power schedule recovered following/ not following LLF rule}
\setlength{\tabcolsep}{2.5pt}
\begin{tabular}{cc|cccccccccccc}
\hline

\multicolumn{2}{c|}{Time step}   & \multicolumn{2}{c}{1}  & \multicolumn{2}{c}{2} & \multicolumn{2}{c}{{\color[HTML]{0000FF} 3}}        & \multicolumn{2}{c}{4}  & \multicolumn{2}{c}{5} & \multicolumn{2}{c}{6} \\
\multicolumn{2}{c|}{P}           & \multicolumn{2}{c}{10} & \multicolumn{2}{c}{0} & \multicolumn{2}{c}{{\color[HTML]{0000FF} 5}}        & \multicolumn{2}{c}{10} & \multicolumn{2}{c}{5} & \multicolumn{2}{c}{0} \\
\multicolumn{2}{c|}{LLF/non-LLF(N)} & LLF      & N     & LLF     & N     & LLF                      & N                  & LLF      & N     & LLF     & N     & LLF     & N     \\ \hline
                             & $e_1$ & 3        & 3           & 2       & 2           & {\color[HTML]{0000FF} 2} & {\color[HTML]{0000FF} 2} & 2        & 1           & 1       & 0           & 0       & 0           \\
                             & $t_{1,e}-t$ & 5        & 5           & 4       & 4           & {\color[HTML]{0000FF} 3} & {\color[HTML]{0000FF} 3} & 2        & 2           & 1       & 1           & 0       & 0           \\
                             & $l_1$ & 2        & 2           & 2       & 2           & {\color[HTML]{0000FF} 1} & {\color[HTML]{0000FF} 1} & 0        & 1           & 0       & 1           & 0       & 0           \\
\multirow{-4}{*}{Request 1}  & $u_1$ & 5        & 5           & 0       & 0           & {\color[HTML]{0000FF} 0} & {\color[HTML]{0000FF} 5} & 5        & 5           & 5       & 0           & 0       & 0           \\ \hline
                             & $e_2$ & 3        & 3           & 2       & 2           & {\color[HTML]{0000FF} 2} & {\color[HTML]{0000FF} 2} & 1        & 2           & 0       & 1           & 0       & 1           \\
                             & $t_{2,e}-t$ & 4        & 4           & 3       & 3           & {\color[HTML]{0000FF} 2} & {\color[HTML]{0000FF} 2} & 1        & 1           & 0       & 0           & 0       & {\color[HTML]{FE0000}-1}           \\
                             & $l_2$ & 2        & 2           & 1       & 1           & {\color[HTML]{0000FF} 0} & {\color[HTML]{0000FF} 0} & 0        & {\color[HTML]{FE0000}-1}          & 0       & {\color[HTML]{FE0000}-1}          & 0       & {\color[HTML]{FE0000}-2}           \\
\multirow{-4}{*}{Request 2}  & $u_2$ & 5        & 5           & 0       & 0           & {\color[HTML]{0000FF} 5} & {\color[HTML]{0000FF} 0} & 5        & 5           & 0       & 0           & 0       & 0           \\ \hline
\end{tabular}
\label{2requestsexample}
\end{table}

\begin{algorithm}[h]
    \caption{Power dispatch with Least-laxity first (LLF) rule}
    \label{A1}
    \begin{algorithmic}[1]
    \REQUIRE Maximum power $\bar{u}$; Initialize HVAC requests $\mathcal{J}=\left( t_{i,s},t_{i,e},\tau_i,l_i \right)$
    \STATE For $t$ in range $[0,T]$
        \STATE Initialize the allocated power budget $B=0$, power rate of each HVAC $u_i[t]=0$;
    \STATE Solve for total power injection from the RL-based controller: $P[t]=\mu(s|\theta^{\mu})$;
    \WHILE{$B<P[t]$ and $\{i \in \mathcal{C}[t] \mid u_{i}[t] = 0\} \neq \varnothing$}
    \STATE Search for the least-laxity HVAC $k=\arg \min _{i:u_i[t]=0}l_i[t]$;
    \STATE Set $|u_k[t]|=\min(\bar{u}, P[t]-B)$ and determine cooling/heating based on $x_k[t]$;
    \STATE $B \leftarrow B+u_k[t]$
    \ENDWHILE
    \STATE $t \leftarrow t+1$, rollout HVAC dynamics.

    \end{algorithmic}
    \end{algorithm}
    
\subsection{Markov Decision Process}
In this subsection, we will model HVAC operation problem as a Markov Decision Process (MDP) denoted as $(\mathcal{S}, \mathcal{A}, \mathcal{P}, \mathcal{R}, \gamma)$, which will be introduced in detail as follows.

\textbf{State.} At each time step $t$,  the state $s[t] \in \mathcal{S}$ consists of the real-time status of all HVAC systems as well as external electricy price signals. As discussed above, the laxity of an HVAC system measures both the remaining duration time and energy demand. It provides information for the controller to determine a feasible total power schedule. Moreover, as we also aim at minimizing energy costs, the external electricity price is necessary for controller to make power dispatch decision. Once receiving current requests $\mathcal{C}[t]$, the state is represented as $s[t]=\left[c[t],\left\{\ell_{i}[t]\right\}_{i \in \mathcal{C}[t]}\right]$.

\textbf{Action.} The action $a[t] \in \mathcal{A}$ for our RL agent is defined as the real-time total power injection $a[t]=P[t]$.

\textbf{Reward.} The reward function $R: \mathcal{S}\times\mathcal{A} \rightarrow \mathbb{R}$ assigns a scalar reward $r$ to guide the agent towards achieving the specified goal. Since HVAC control objectives involve controlling indoor temperature within an expected duration time and minimizing energy costs, the reward $r[t]$ is composed of two terms: the laxity part $r_1[t]=\sum_{i\in\mathcal{C}[t]}\ell_i[t]$;  and the total energy cost: $r_2[t]=-c[t] P[t]$ with weighting parameters ($\alpha, \beta$):

\begin{equation}
    r[t]=\alpha r_1[t]+\beta r_2[t].
\end{equation}


\textbf{Transition.} The transition function $\mathcal{P}: \mathcal{S}\times\mathcal{A}\rightarrow\mathcal{S}$ returns the next state given current state-action pair. At timestep $t_{i,s}$,  the state of each HVAC system is initialized as a new request $e_{i}[t]$ calculated by the cooling/heating time defined in \eqref{equ:request}, and laxity calculated as \eqref{equ:laxity}.
Given an HVAC request, with current $\ell_i[t]$ and temperature $x_i[t]$, state transition is given by:

\begin{subequations}
\begin{align}
    x_i[t+1] &= e^{-a \Delta t}(x_i[t]-\frac{b u_i[t]}{a}-x_{out}[t])+\frac{b u_i[t]}{a}+x_{out}[t];\\
    e_{i}[t+1]&=\begin{cases}
              &\zeta(x_{i,r},x_i[t+1],\bar{u}_{i},\lambda_i) \quad \text{if $x_i[t+1]< x_{i,r}$}\\
              &0, \quad \text{if $x_i[t+1]= x_{i,r}$}\\
              &\zeta(x_{i,r},x_i[t+1],\underline{u}_{i},\lambda_i) \quad \text{if $x_i[t+1]>x_{i,r};$}\\
              \end{cases}\\
    \tau_{i}[t+1]&=\begin{cases}
                 &\zeta(\underline{x}_{i},x_i[t+1],\bar{u},\lambda_i),  \quad \text{if $x_i[t+1]<\underline{x}_{i}$}\\
                 &0, \quad \text{if $\underline{x}_{i}\leq x_i[t+1]\leq \bar{x}_{i}$}\\
                 &\zeta(\bar{x}_{i},x_i[t+1],\underline{u},\lambda_i),  \quad \text{if $x_i[t+1]>\bar{x}_{i}$};
                 \end{cases}\\
    \ell_i[t+1]&\resizebox{0.87\linewidth}{!}{$=\begin{cases}
               &t_{i,e}-(t+1)-e_i[t+1],  \quad \text{if } t_{i,s}\leq t+1 \leq t_{i,e} \land \underline{x}_{i}\leq x_i[t+1]\leq \bar{x}_{i} \\
               &-\tau_i[t+1],  \quad \quad \quad \quad \quad \quad \quad  \text{if } x_i[t+1]\leq \underline{x}_{i} \lor x_i[t+1]\geq \bar{x}_{i}.
               \end{cases}$}
\end{align}
\end{subequations}

\subsection{State Abstraction}
With multiple HVAC systems involved, the size of state space will become cumbersome, which will cause difficulties for standard RL training. State abstraction helps map the original states to an abstract state with a smaller state space~\cite{jiang2018notes}, while an appropriate abstraction must preserve the necessary information of the original MDP. In this subsection, we give the definition of model-irrelevance abstraction and approximate model-irrelevant abstraction~\cite{li2006towards}, and show how our HVAC state abstraction method holds such conditions.

\textbf{Definition 6 (Model-irrelevance abstraction)} Given an MDP $(\mathcal{S}, \mathcal{A}, \mathcal{P}, \mathcal{R}, \gamma)$, and any two states $s_1, s_2 \in \mathcal{S}$, a model-irrelevance abstraction $\phi_{model}$ is such that for any action $a\in \mathcal{A}$ and any abstract next-step state $\hat{s}\in\phi_{model}(\mathcal{S})$, $\phi_{model }\left(s_1\right)=\phi_{model}\left(s_2\right)$ implies:
\begin{equation}
\label{equreward}
    r\left(s_1, a\right)=r\left(s_2, a\right);
\end{equation}
\begin{equation}
\label{equtran}
    P\left(\hat{s} \mid s_1, a\right)= P\left(\hat{s} \mid s_2, a\right).
\end{equation}

With model-irrelevance abstraction, $\mathcal{S}$ and $\phi_{model}(\mathcal{S})$ attain the same value functions and thus the same optimal policies for any given action. However, model-irrelevance abstraction require strict conditions, while exact abstractions are hard to find and verify in practice. Thus, we consider approximate model-irrelevant abstraction for our LLF-based approach.

\textbf{Definition 7 ($\left(\epsilon_R, \epsilon_P\right)$-approximate model-irrelevant abstraction)} Given an MDP $(\mathcal{S}, \mathcal{A}, \mathcal{P}, \mathcal{R}, \gamma)$, and any states $s_1, s_2 \in \mathcal{S}$, a $\left(\epsilon_R, \epsilon_P\right)$-approximate model-irrelevant abstraction $\phi_{model}$ is such that for any action $a\in \mathcal{A}$ and any abstract state $\hat{s}\in\phi_{model}(\mathcal{S})$, $\phi_{model }\left(s_1\right)=\phi_{model}\left(s_2\right)$ implies:
\begin{equation}
\label{equreward}
    \left|r\left(s_1, a\right)-r\left(s_2, a\right)\right| \leq \epsilon_R
\end{equation}
\begin{equation}
\label{equtran}
    \left|P\left(\hat{s} \mid s_1, a\right)-P\left(\hat{s} \mid s_2, a\right)\right| \leq \epsilon_P
\end{equation}

In this paper, given original state $s[t]=\left[c[t],\left\{\ell_{i}[t]\right\}_{i \in \mathcal{C}[t]}\right]$, we redefine the state space in MDP with approximate model-irrelevance abstraction:


\textbf{Abstract State.} The abstract state $\hat{s} \in \hat{\mathcal{S}}$ is defined as $\hat{s}[t] = \phi_{model}(s[t])=\left[c[t], L[t]\right]$, where $L[t]=\sum_{i \in \mathcal{C}[t]} \ell_i[t]$.

To show equivalent approximate model-irrelevance abstraction, we assume there are two original states $s_1[t]=\left[c_1[t],\left\{\ell_{i}[t] \right\}_{i \in \mathcal{C}_1}\right]$ and $s_2[t]=\left[c_2[t],\left\{\ell_{j}[t]\right\}_{j \in \mathcal{C}_2}\right]$ that can be mapped to the same state $\hat{s}$. Since $\phi_{model}(s_1[t])=\phi_{model}(s_2[t])$, we have $c_1[t]=c_2[t]$ and $L_1[t]=L_2[t]$. 

It is obvious that given the same action $a=P[t]$, $R\left(s_{1}, a\right) = -c[t]P[t]+L_1[t] = -c[t]P[t]+L_2[t] = R\left(s_{2}, a\right)$.

As for condition \eqref{equtran}, it is possible to quantify the distance between laxity term in $s_1$ and $s_2$. Assume $k=\min(\frac{P[t]}{\bar{u}},N)$. For any state that can be mapped to $\hat{s}$, we divide the HVAC systems into two sets based on whether they are chosen to be operated by the LLF rule (set $\mathcal{C}$) or not (set $\mathcal{C}^{'}$). 
As shown in \textbf{Proposition 2}, for $i\in\mathcal{C}$, with $|u_i[t]|=\bar{u}$, $\ell_i[t+1]-\ell_i[t]=-1+(e_i[t]-e_i[t+1])=-1-\frac{1}{a_i \Delta t} \ln \left(\frac{e^{-a_i \Delta t}(x_i[t]-\frac{b_i \bar{u}}{a_i}-x_{out}[t])+\frac{b_i \bar{u}}{a_i}-\frac{b_i \bar{u}}{a_i}}{x_i[t]-\frac{b_i\bar{u}}{a_i}-x_{out}[t]}\right)=0,\forall x_i[t]<x_{i,r}; \ell_i[t+1]-\ell_i[t]=-1+(e_i[t]-e_i[t+1])=-1-\frac{1}{a_i \Delta t} \ln \left(\frac{e^{-a_i \Delta t}(x_i[t]-\frac{b_i \underline{u}}{a_i}-x_{out}[t])+\frac{b_i \underline{u}}{a_i}-\frac{b_i \underline{u}}{a_i}}{x_i[t]-\frac{b_i\underline{u}}{a_i}-x_{out}[t]}\right)=0,\forall x_i[t]>x_{i,r} $. For $i\in\mathcal{C}^{'}$, with $u_i[t]=0$, $\ell_i[t+1]-\ell_i[t]=-1+(e_i[t]-e_i[t+1])=-1-\frac{1}{a_i \Delta t} \ln \left(\frac{e^{-a_i \Delta t}(x_i[t]-x_{out}[t])-\frac{b_i \bar{u}}{a_i}}{x_i[t]-\frac{b_i\bar{u}}{a_i}-x_{out}[t]}\right).$ The aggregated laxity in next time step is given as $L[t+1]=L[t]-(n-\frac{P[t]}{\bar{u}}) + \sum_{i\in\mathcal{C}^{'}}(e_i[t]-e_i[t+1]) = L[t]-(n-\frac{P[t]}{\bar{u}}) - \sum_{i\in\mathcal{C}^{'}}\left[\frac{1}{a_i \Delta t} \ln \left(\frac{e^{-a_i \Delta t}(x_i[t]-x_{out}[t])-\frac{b_i \bar{u}}{a_i}}{x_i[t]-\frac{b_i\bar{u}}{a_i}-x_{out}[t]}\right)\right].$ Hence, given two original state $s_1[t]$ and $s_2[t]$ that can be mapped to the same state $\hat{s}[t]$ as well as action $P[t]$,  
the deviation between the next step's state $\hat{s}_1[t+1]$ and $\hat{s}_2[t+1]$ is only related to aggregated laxity difference, which is a bounded difference given the total number of requests and timestep $\Delta t$: 
\begin{equation}
\begin{aligned}
    &L_1[t+1]-L_2[t+1]\\
    &=\sum_{i\in\mathcal{C}_2^{'}}\left[\frac{1}{a_i \Delta t} \ln \left(\frac{e^{-a_i \Delta t}(x_i[t]-x_{out}[t])-\frac{b_i \bar{u}}{a_i}}{x_i[t]-\frac{b_i\bar{u}}{a_i}-x_{out}[t]}\right)\right]\\
    &-\sum_{i\in\mathcal{C}_1^{'}}\left[\frac{1}{a_i \Delta t} \ln \left(\frac{e^{-a_i \Delta t}(x_i[t]-x_{out}[t])-\frac{b_i \bar{u}}{a_i}}{x_i[t]-\frac{b_i\bar{u}}{a_i}-x_{out}[t]}\right)\right].
\end{aligned}
\end{equation}

Then it is sufficient to train an RL agent only on aggregated laxity $L[t]$, while the resulting policy based on RL and LLF rule is approximate equivalent to standard RL training paradigm using full states.

\subsection{RL with State Abstraction}
After state abstraction, the new state space becomes $\hat{\mathcal{S}}=\left[c[t], L[t]\right]$. As illustrated in Fig.\ref{framework1}, the role of the controller is to receive the aggregated states as well as the price signal from the electricity market, and then schedule the total power at the current time step. Under such a framework, the controller only knows the state after abstraction from the aggregator, but not the physical model nor the information of individual HVAC systems. In addition, both state space and action space are continuous. As a result, here we propose to use deep deterministic policy gradient (DDPG) algorithm to efficienct learn the controller~\cite{lillicrap2015continuous}. DDPG is a  model-free reinforcement learning algorithm tailored for continuous action space, and it adopts the actor-critic framework, in which the actor network $\mu(\hat{s}|\theta^{\mu})$ specifies the current policy   mapping states to an action deterministically, while the critic network $Q(\hat{s},a|\theta^{Q})$ is the approximated action-value function learned using Bellman equation, where $\theta$ denotes the parameters of neural networks. Target critic network $Q^{\prime}(\hat{s},a|\theta^{Q^{\prime}})$ and target actor network $\mu^{\prime}(\hat{s}|\theta^{\mu^{\prime}})$ are utilized to stabilize the parameter updating.

To be specific, the action-state value function describes the expected return when taking action $a[t]$ given state $\hat{s}[t]$: 
\begin{equation}
    Q(\hat{s}[t],a[t]) = \mathbb{E}\{\sum_{k=0}^{\infty}\gamma^{k}r[t+k+1](\hat{s}[t],a[t])\}.
\end{equation}
The critic network is optimized by minimizing its loss with target value function:
\begin{equation}
    \mathcal{L}\left(\theta^Q\right)=\mathbb{E}_{\hat{s}[t] \sim \rho^\mu, a[t] \sim \mu, r[t] \sim E}\left[\left(Q\left(\hat{s}[t], a[t] \mid \theta^Q\right)-y[t]\right)^2\right],
\end{equation}
where $y[t]$ is the target value based on target critic network and traget actor network:
\begin{equation}
    y[t]=r[t]+\gamma Q^{\prime}\left(\hat{s}[t+1], \mu^{\prime}\left(\hat{s}[t+1] \mid \theta^{\mu^{\prime}}\right) \mid \theta^{Q^{\prime}}\right).
\end{equation}
The actor network is updated by policy gradient approach:
\begin{equation}
\begin{aligned}
\nabla_{\theta^\mu} J & \approx \mathbb{E}_{\hat{s}[t] \sim \rho^\mu}\left[\left.\nabla_{\theta^\mu} Q\left(\hat{s}, a \mid \theta^Q\right)\right|_{\hat{s}=\hat{s}[t], a=\mu\left(\hat{s}[t] \mid \theta^\mu\right)}\right] \\
&\resizebox{0.8\linewidth}{!}{$=\mathbb{E}_{\hat{s}[t] \sim \rho^\mu}\left[\left.\left.\nabla_a Q\left(\hat{s}, a \mid \theta^Q\right)\right|_{\hat{s}=\hat{s}[t], a=\mu\left(\hat{s}[t]\right)} \nabla_{\theta_\mu} \mu\left(\hat{s} \mid \theta^\mu\right)\right|_{\hat{s}=\hat{s}[t]}\right]$}.
\end{aligned}
\end{equation}

\section{Case Study}
\label{case study}
\begin{figure}[htbp]
\centerline{\includegraphics[width=0.45\textwidth]{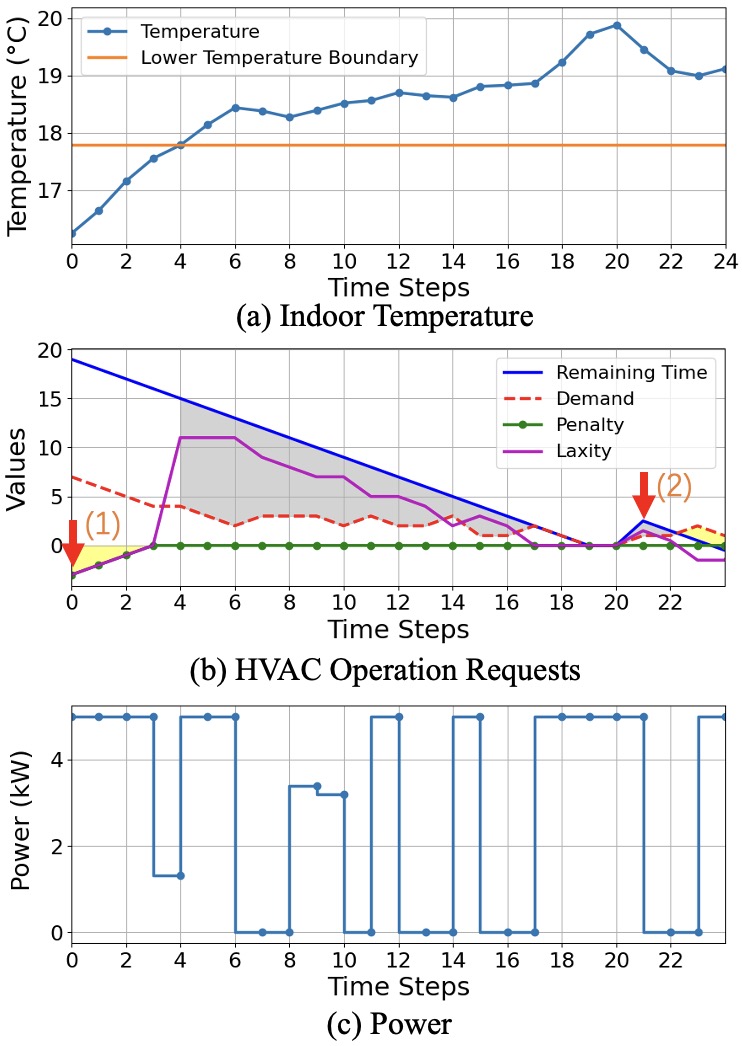}}
\caption{Simulation result of an HVAC system on May $1^{st}$, 2015 (case 2). (a) Real-time outdoor temperature. (b) At the $0^{th}$ time step and the $21^{st}$ time step, this HVAC system sends operation requests respectively. Yellow shaded area denotes negative laxity due to comfort range violations. During the $4^{th}$ and $21^{st}$ time steps, the value of laxity equals the remaining operation time minus minimum heating/cooling time, which equals the height of the grey shaded area. After the $22^{nd}$ step, the remaining time is not enough for the remaining energy demand, thus the laxity is negative again, which equals the height of the yellow shaded area. (c) The power dispatched to this HVAC system based on LLF rules.}
\label{laxity}
\end{figure}

\subsection{Experiment Setup}
We use real-world hourly dataset \cite{data} for algorithm validations. This dataset contains Spain's 4 years of electrical consumption, pricing from Spanish TSO Red Electric España, and weather data. Each training and testing episode contains 96 timesteps, and the time interval is 1 hour. To test the experiment performance in different weather situations, we show 3 typical cases: (1) Feb. 1-4, 2015 (case 1): the HVAC systems need to control heating; (2) May. 1-4, 2015 (case 2): the HVAC system should both heat and cool; (3) Aug. 1-4, 2015 (case 3): the HVAC system needs to cool to keep a comfortable temperature. The RL simulation environment consists of temperature dynamics models for both single-zone and multi-zone commercial building clusters. Specifically, the single-zone environment comprises 10 individual commercial buildings, each equipped with an HVAC system that can send operation requests to the aggregator. At each time step, the buildings update their laxity and receive a dispatched power signal from the aggregator, and then the state (e.g., indoor temperature) evolves based on received power dispatch actions and individual dynamics. The multi-zone environment includes 10 multi-zone commercial buildings~(see Fig. \ref{RC}), and each zone is equipped with an HVAC system. Each zone updates and reports its zonal laxity directly to the aggregator and receives a dispatched power signal from the aggregator. Our RL agent utilizes aggregated  laxity $L(t)$ from all zones, and  LLF rule still applies to each single zone. The whole building updates its states based on building-level dyanmics\eqref{state function}. 

\begin{figure}[htbp]
\centerline{\includegraphics[width=0.4\textwidth]{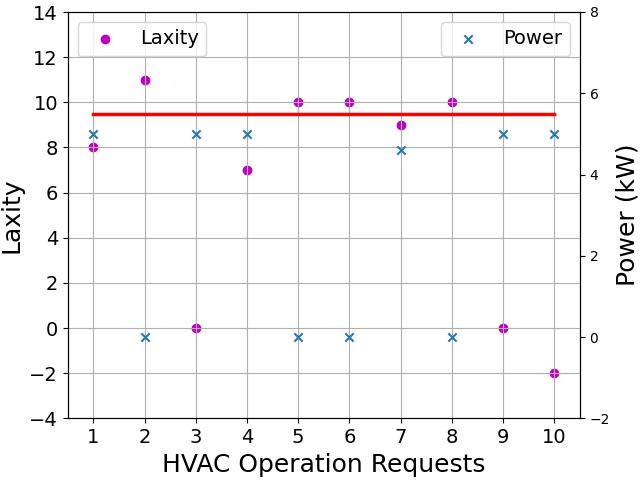}}
\caption{Laxity representation and power dispatch on 10 HVAC systems. Given total power dispatch from RL controller, the six HVAC operation requests with the smallest laxity value (lower than the red line) are chosen to be operated.}
\label{10requests}
\end{figure}

\begin{figure}[htbp]
\centerline{\includegraphics[width=0.5\textwidth]{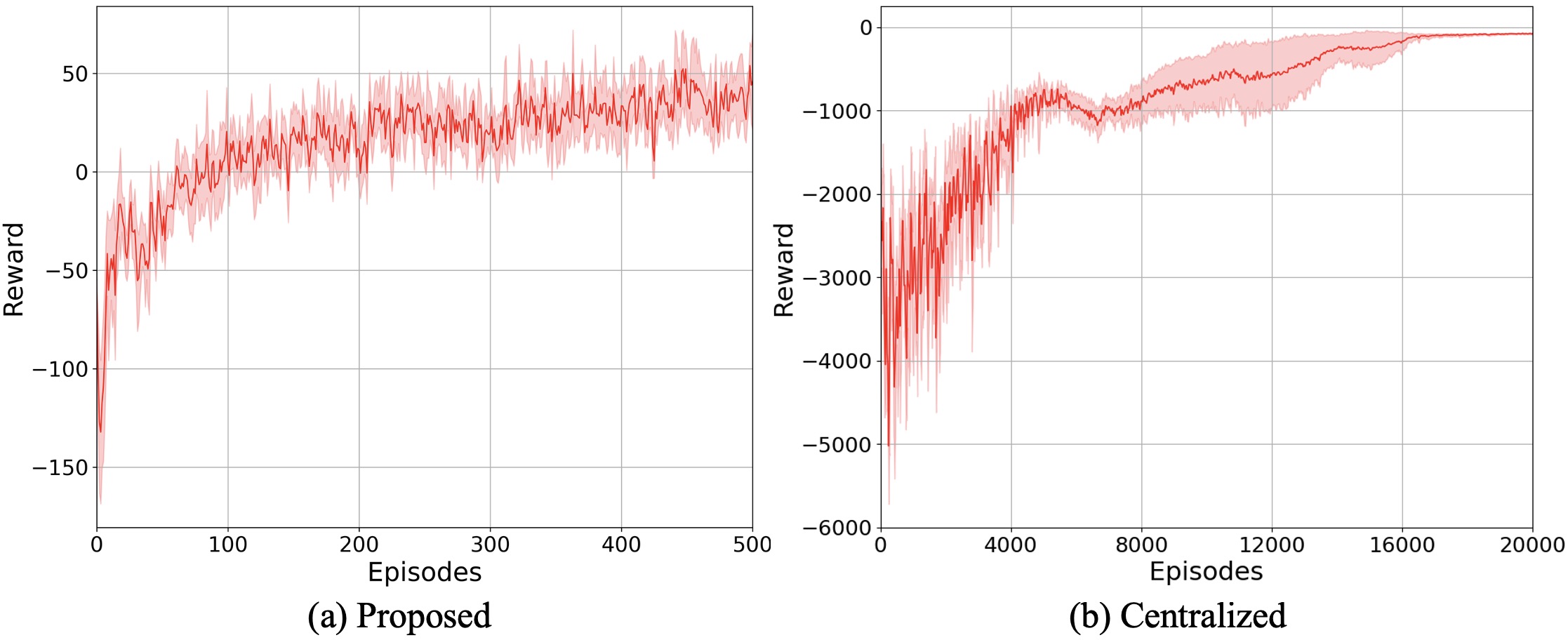}}
\caption{Training curves of \texttt{Proposed} and \texttt{Centralized} algorithms, showing the episode reward.}
\label{reward}
\end{figure}

We compare our proposed method (\texttt{Proposed}) with two benchmark methods:\\
1) \emph{Model Predictive Control (\texttt{MPC}):} The MPC model has access to the exact HVAC dynamics and full information on electricity price, which can find the optimal HVAC schedules and is considered as an optimal benchmark. 
The objective of MPC optimization problem is minimizing energy costs subject to target temperature deviation and power limits.\\
2) \emph{Centralized RL (\texttt{Centralized}):} This scheme directly uses Proximal Policy Optimization (PPO) to control the power of all HVAC systems. To be specific, the states include the temperature of each zone, outdoor temperature, and electricity price; the actions are the power of each HVAC system; the reward function is related to the temperature difference between the target temperature and current temperature of each zone; and the transition function only updates temperature of each zone without laxity.

To measure temperature control and energy cost saving performance, we introduce two performance metrics: \\
1) \emph{Average temperature deviation (ATD):} \\$ATD=\frac{1}{TN}\sum_{t \in \mathcal{T}}\sum_{i \in \mathcal{N}}\left|x_{i,t}-x_{i,target}\right|$;\\
2) \emph{Total Energy Cost (TEC):} $TEC=\sum_{t \in \mathcal{T}}\sum_{i \in \mathcal{N}}p(t)P(t)$.

\begin{figure*}[htbp]
\centerline{\includegraphics[width=0.9\textwidth]{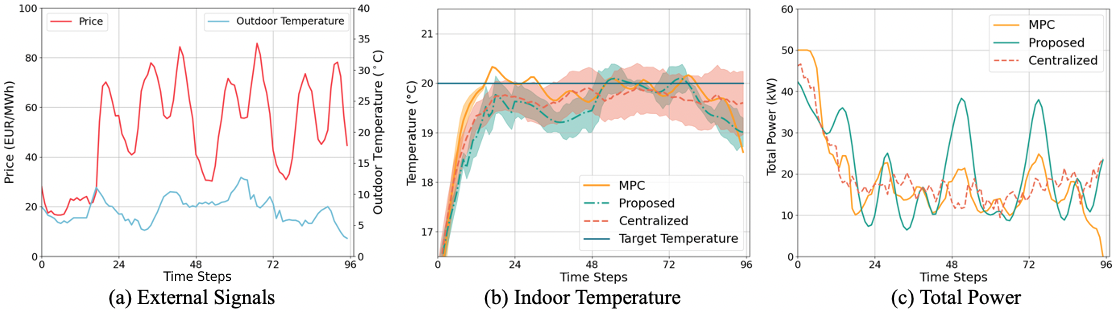}}
\caption{(a) Real-time external signals, (b) temperature control results, and (c) total power scheduled by controllers on Feb. 1-4, 2015 (case 1).}
\label{case1}
\end{figure*}

\begin{figure*}[htbp]
\centerline{\includegraphics[width=0.9\textwidth]{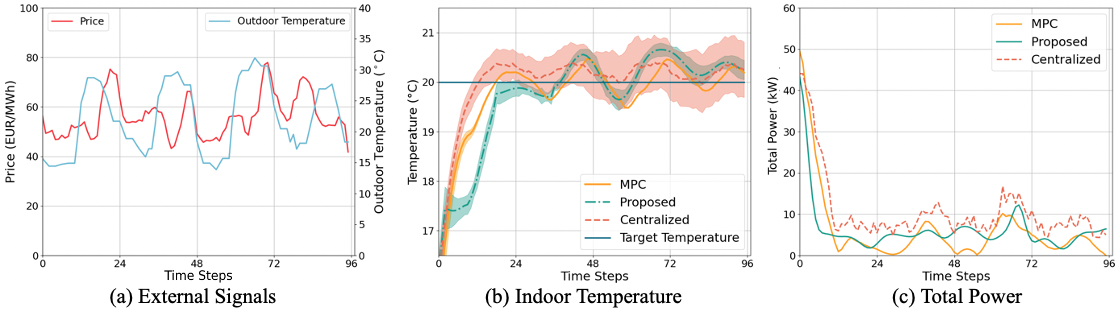}}
\caption{(a) Real-time external signals, (b) temperature control results, and (c) total power scheduled by controllers on May. 1-4, 2015 (case 2).}
\label{case2}
\end{figure*}

\begin{figure*}[htbp]
\centerline{\includegraphics[width=0.9\textwidth]{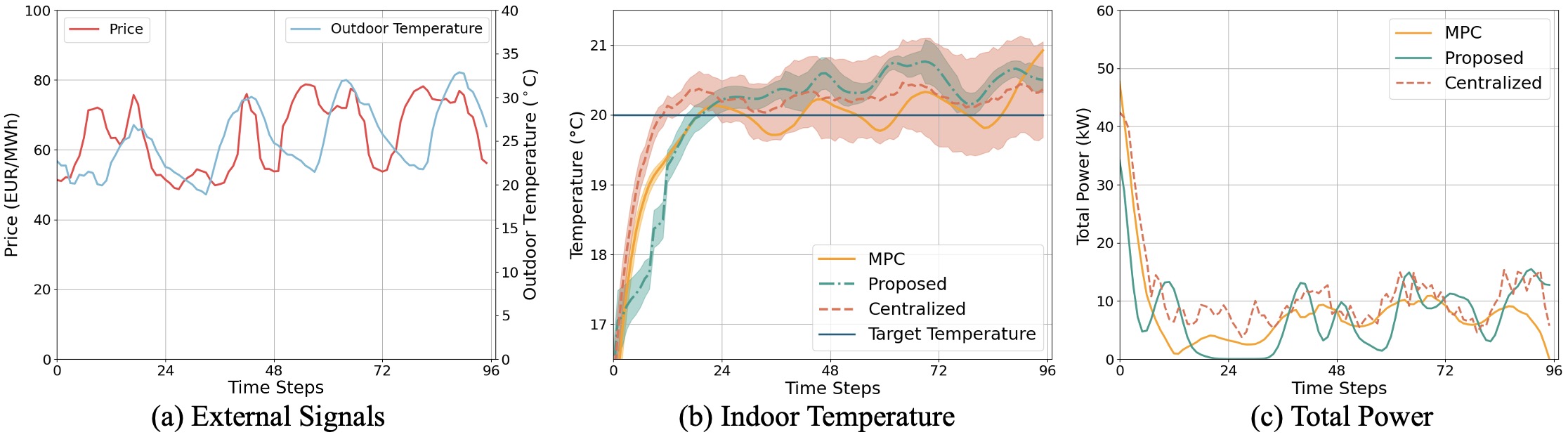}}
\caption{(a) Real-time external signals, (b) temperature control results, and (c) total power scheduled by controllers on Aug. 1-4, 2015 (case 3).}
\label{case3}
\end{figure*}

\subsection{Laxity Representation Analysis}
\label{Laxity Representation Analysis}

Fig. \ref{laxity} presents the relationship between temperature, laxity, and dispatched power of an HVAC system on May 1, 2015 (case 2). At the $0^{th}$ and $21^{th}$ time steps, this HVAC system sends operation requests. The yellow shaded area denotes two anomalous situations where laxity becomes negative. It is noteworthy that during the $0^{th}$ to $4^{th}$ time steps, the system experiences an abnormal condition due to the temperature falling below the lower boundary, thus the laxity value is equal to the negative penalty value as calculated in equation \eqref{augment}. Another abnormal situation appears when the minimum heating/cooling time is larger than the remaining operation time, indicating that the operation request can not be finished within the expected duration time. In the $23^{th}$ time step, the minimum heating/cooling time is larger than the remaining duration time, thus the laxity value is negative, and the request can not be satisfied before the remaining duration time reaches zero. When an abnormal situation appears, the laxity value is negative, and according to the LLF rules, these requests can be operated with higher priority. It can also be observed that in most of the time steps, the laxity is decreasing, as discussed in \textbf{Proposition 2}. 
Note that the analysis \textbf{Proposition 2} assumes that the outdoor temperature keeps almost unchanged in two adjacent time steps, while the actual laxity value may infrequently increase.

In order to illustrate the power dispatching process based on the constraint-augmented LLF rule, we plot a single-step laxity and power dispatching result for a group of 10 HVAC systems in May $1^{st}$, 2015 (case 2). In this case, the total power scheduled by the controller is $29.608$, according to \ref{A1}, the 6 HVAC operation requests with the lowest laxity values are chosen to be operated.

\subsection{Performance Analysis}
Fig. \ref{reward} presents the average and standard deviation of the episodic training reward for \texttt{Proposed} and \texttt{Centralized} across different numbers of episodes, revealing their convergence speeds. 
The number of states increases with the number of HVAC systems in \texttt{Centralized}, leading to a much slower convergence speed. 

In Table. \ref{results}, we measure the ATD and TEC metrics for all settings. It can be seen that \texttt{Proposed}  achieves smaller ATD and TEC than \texttt{Centralized} in all three cases, which indicates that \texttt{Proposed} has better performance in temperature control and energy cost saving in different situations. Moreover, with similar ATD values, since the TEC value of \texttt{Proposed} is much lower than that of \texttt{Centralized}, this also suggests that \texttt{Proposed} has the potential to change the temperature and energy saving performance by further adjusting the weight factor of the reward function in reinforcement learning according to the actual needs and preferences of users.

Fig. \ref{case1} to Fig. \ref{case3} illustrate the average temperature, total power based on baseline approaches and \texttt{Proposed}, and external signals in three cases. In Fig. \ref{case1}(b) Fig. \ref{case2}(b), and Fig. \ref{case3}(b), we plot the average temperature curve with standard deviation of 10 buildings. It is evident that the temperature deviation of 10 HVAC systems is significantly lower when operated under \texttt{Proposed} compared to \texttt{Centralized}, thereby demonstrating the superior temperature control performance of the \texttt{Proposed} among a group of HVAC systems. Importantly, the \texttt{Proposed} involves three steps - aggregating state information using an aggregator, determining the total power by the controller, and scheduling power by the aggregator. The effective performance of these three steps in the algorithm is evidenced by the superior temperature control demonstrated. 

Fig. \ref{case1}(c), Fig. \ref{case2}(c), and Fig. \ref{case3}(c) further illustrate \texttt{Proposed} can achieve significantly smaller energy cost (TEC) compared to the centralized algorithm. From the graphs, we can observe that \texttt{Proposed} implements a controller which can respond to external price and temperature signals to control the total power. For instance, as shown in Fig. \ref{case1}(a), during the $24^{th}$, $48^{th}$, and $72^{th}$ time steps, there are dips in the price, and the controller accordingly increases the total power in Fig. \ref{case1}(c) to complete the HVAC operation requests with lower energy costs. Note that although our controller cannot directly access temperature data, external temperatures affect the laxity, which in turn affects the input of the controller. For example, as shown in Fig. \ref{case3}, in the $42^{th}$, $64^{th}$, and the $88^{th}$ time steps, the outdoor temperature peaks in Fig. \ref{case3}(a) correspond to the total power peak in Fig. \ref{case3}(c) to meet the cooling demand.

\begin{table}[htbp]
\caption{Comparison of experiment results.}
\begin{tabular}{l|cccccc}
\hline
\multicolumn{1}{c|}{\multirow{2}{*}{Methods}} & \multicolumn{2}{c}{Case 1} & \multicolumn{2}{c}{Case 2} & \multicolumn{2}{c}{Case 3} \\
\multicolumn{1}{c|}{}                         & ATD         & TEC          & ATD         & TEC          & ATD         & TEC          \\ \hline
MPC                                           & 0.452       & 75.422       & 0.466       & 31.470       & 0.431       & 47.671       \\
Proposed                                      & 0.814       & 79.626       & 0.823       & 32.610       & 0.799       & 46.826       \\
Centralized                                   & 0.945       & 85.739       & 0.823       & 56.356       & 0.872       & 66.418       \\ \hline
\end{tabular}
\label{results}
\end{table}

\subsection{Scalability Analysis}
\begin{figure}[h]
\centerline{\includegraphics[width=0.5\textwidth]{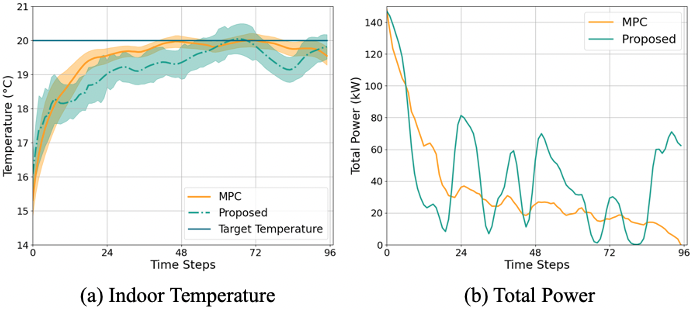}}
\caption{Simulation results of the multi-zone scenario.}
\label{multi-zone}
\end{figure}

We conduct the multi-zone building energy management to validate algorithm's scalability. 
As the number of zones increases and the system's thermal dynamics becomes more complex, the input states of the RL controller do not increase with the increase of zones, and only keep the electricity price along with aggregated laxity . As \texttt{Centralized} cannot train a convergent RL policy, it suggests that with the expansion of the state space, \texttt{Centralized} cannot maintain the temperature control performance. Therefore, we only compared the performance of \texttt{MPC} and \texttt{Proposed}. From Fig. \ref{multi-zone}(a), it can be observed that based \texttt{Proposed}, the HVAC system is still able to adjust the indoor average temperature from the initial temperature to near the target temperature and maintain it there. As for total power, we observed in From Fig. \ref{multi-zone}(c) that several peaks in the total power curve are related to variations in the price and outdoor temperature. Although the energy-saving effect is slightly worse than that of the model-based \texttt{MPC}, our model-free control method can be applied to large-scale and complex thermodynamic systems.

\begin{figure}[h]
\centerline{\includegraphics[width=0.5\textwidth]{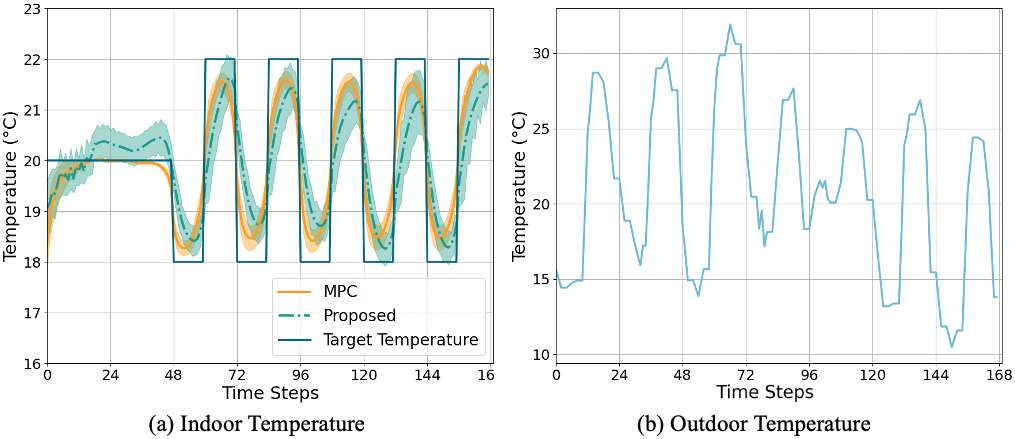}}
\caption{Simulation results of the week-long scenario.}
\label{7days}
\end{figure}

We further simulate a week-long real-world scenario to test the effectiveness of our algorithm. During this week, we assume that the temperature control requirements of the building users may vary depending on different preferences, such as day and night, weekdays and weekends. The target temperature curve in Fig. \ref{7days}(a) illustrates the temperature control target in detail. The results demonstrate that \texttt{Proposed} can appropriately control the total power and dispatch power for HVAC systems to track the input target temperature signal.

\section{Conclusion}
\label{conclusion}

In this work, we propose a laxity-aware reinforcement learning approach for efficient control of multiple HVAC systems. The proposed model-irrelevant state abstraction method can seamlessly aggregate laxity information and disaggregate power dispatch by training a laxity-aware RL agent. With extensive numerical studies on real-world data, our proposed framework can accurately represent the emergency level of an HVAC operation request under different situations, and effectively dispatch power for the requests using LLF rules. Moreover, the proposed framework can achieve better temperature control and energy cost savings than centralized learning-based method. 
Multi-zone and week-long simulations further indicate scalability and generality of our proposed method. In the future work, we will explore the model-free control with laxity representation for grid-connected systems. We are also interested in designing more general flexibility aggregation schemes using laxity information.


\bibliographystyle{IEEEtran}
\bibliography{bib}
\end{document}